\documentclass[11pt]{article}
\usepackage{graphicx,moriond,epsfig}

\begin{document}
\vspace*{4cm}
\title{VIOLENT STAR AND STAR CLUSTER FORMATION IN NEARBY AND DISTANT GALAXIES}

\author{ UTA FRITZE -- v. ALVENSLEBEN}
\address{Institut f\"ur Astrophysik, Universit\"at G\"ottingen, Friedrich-Hund-Platz 1, 37077 G\"ottingen, Germany}

\maketitle
\abstracts{I present recent observations and analyses of star cluster formation 
in a wide variety of environments -- from young star clusters and super star 
clusters in normal actively star-forming spirals and irregulars to starbursting 
dwarfs and spiral-spiral mergers. Star cluster formation in interacting 
galaxies can be restricted to central starburst region, extend over the entire body of the merger, or even all along extended 
tidal structures. I address methods and results for the determination of star 
cluster ages, metallicities, masses, and sizes and discuss the nature, 
possible lifetimes and future signatures of these star cluster populations, 
as well as the relative importance of field star formation vs. star cluster 
formation.}

\noindent
{\small¥{\it Keywords}: Stars: formation, Galaxies: evolution, formation, 
interactions, ISM, starburst, star clusters, Globular Clusters: general, 
Open Clusters and Associations: general}

\section{Introduction}
I briefly and selectively review some aspects of 
{\bf S}tar {\bf F}ormation ({\bf SF}) in the context of a comparison between normal  
SF and violent SF in starbursting and interacting galaxies and between SF in the
local universe and at high redshift. {\bf S}tar {\bf C}luster 
({\bf SC}) formation is an important mode of SF, in particular during violent 
SF episodes. SCs -- as far as they survive -- are much better tracers of 
(violent) 
SF in galaxies than integrated light because they can be studied one by one. 
The age and metallicity distributions of SC and {\bf G}lobular {\bf C}luster 
({\bf GC}) systems hold unique clues about the formation histories of their
parent galaxies over cosmological timescales. 

In Sect. 2. I will review some aspects of normal vs. violent SF, in Sect. 3. I
sketch the present state of our knowledge about SC formation, both in terms of 
observational evidence and theoretical concepts, and in Sect. 4. I discuss the
relation between SC formation and field star formation and in Sect. 5 I summarise and present an
embarrassingly long list of open issues. 

\section{Star Formation : normal and violent}
{\bf S}tar {\bf F}ormation {\bf R}ates ({\bf SFR}s) in nearby normal and
starburst galaxies are conventionally derived from their
H$_\alpha$-luminosities via 
$${\rm SFR ~[M_{\odot}/yr] ~=~ L(H_{\alpha}~/~1.26 \cdot 10^{41})~[erg/s].}$$
This assumes a Salpeter IMF from ${\rm 0.1 - 100~M_{\odot}}$ and approximately
solar metallicity (cf. e.g. Kennicutt 1998). As we have shown in Weilbacher \&
Fritze -- v. Alvensleben (2001) using our GALEV evolutionary synthesis models, this relation is 
only valid as long as SFRs do not fluctuate on
timescales $\leq 10^7$ yr. In the case of individual SFing regions or
for starbursting dwarf galaxies which have SF fluctuations on timescales of
$10^5$ to $10^6$ yr, the SFRs estimated from their H$_\alpha$-luminosities can
be wrong by as much as a factor of $\sim 100$, because changes in H$_\alpha$
emission lag behind changes in the SFR by about the lifetime of the most massive
stars. Moreover, because low metallicity
stellar populations are brighter and have much stronger ionising fluxes than
solar metallicity ones, the above relation becomes metallicity dependent. For
${\rm Z=1/20 \cdot Z_{\odot}}$, e.g., SFRs derived from the above relation are 
overestimated by a factor $\leq 3$ for continuous SF and by a factor $\geq 3$
for starbursts. 

\smallskip\noindent
Tight correlations are observed between H$_\alpha$-derived SFRs and UV-, mid-IR-,
FIR-, and radio-luminosities, that then, in turn, can also be used to estimate
galaxy SFRs. 

\smallskip\noindent
For distant galaxies, SFRs are often derived from their [OII]3727-luminosities
via 
$${\rm SFR ~[M_{\odot}/yr] ~=~ L([OII]~/~7.14 \cdot 10^{40})~[erg/s].}$$ 
The metallicity dependence of the [OII]3727$-$line is twofold.
[OII] fluxes depend on the oxyen abundance and, hence, increase with increasing 
metallicity of the ionised gas. They also depend on the strength of the ionising flux that
decreases with increasing metallicity. The combination of both effects 
accounts for 
a factor $\sim 2$ change from high to low metallicity in the transformation factor between ${\rm L([OII])}$ 
and SFR (see Weilbacher \& Fritze -- v. Alvensleben 2001), resulting in an overestimate of SFRs based
on [OII] in low metallicity galaxies when using the conventional relation
derived for local near-solar metallicity galaxies (cf. Kewley {\sl et al.} 2004,
Bicker \& Fritze -- v. Alvensleben {\sl submitted}).  

\noindent
SFRs are of order ${\rm 1 - 3~M_{\odot}/yr}$ for spiral galaxies with masses 
around ${\rm 10^{10}~M_{\odot}}$ and of order ${\rm 0.01 -
3~M_{\odot}/yr}$ for irregular and dwarf irregular galaxies with masses in 
the range ${\rm 10^6~to~10^9~M_{\odot}}$. Starbursts in isolated dwarf galaxies
have SFRs of order ${\rm 0.1 -
10~M_{\odot}/yr}$.  

\noindent
Bursts strengths -- defined as the relative increase of the stellar mass during 
a burst ${\rm b:=\Delta S_{burst} / S}$ can reasonably be derived for young
post-starbursts only. For a sample of BCDGs with optical
and NIR photometry burst strengths have been shown 
by Kr\"uger, Fritze -- v. Alvensleben \& Loose (1995) to range from ${\rm b=0.001}$ to
${\rm b=0.05}$, 
and to decrease with increasing total mass, including their
important HI masses, 
in agreement with expectations from stochastic self-propagating SF scenarios (cf. Fig.1). 

\begin{figure}
\begin{center}
\psfig{figure=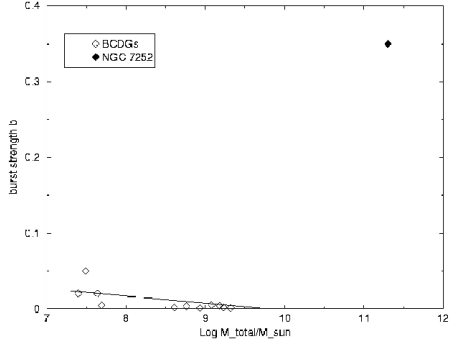,height=6.cm,angle=-90.}
\end{center}
\caption{Burst strengths vs. total galaxy mass for the sample of BCDGs from 
Kr\"uger, Fritze -- v. Alvensleben \& Loose (1995) and for the post-starburst galaxy NGC 7252, for which the symbol only gives a lower limit (Fritze -- v. Alvensleben \& Gerhard 1994b).}
\end{figure}

Massive gas-rich interacting galaxies feature high and sometimes very high SFRs
of order $50,~100,$ up to ${\rm 1000~M_{\odot}/yr}$ and more for {\bf L}uminous and
{\bf U}ltraluminous {\bf IR} {\bf G}alaxies ({\bf LIRG}s and {\bf ULIRG}s) and
their higher redshift counterparts, the SCUBA galaxies, in
their global or nuclear starbursts which typically last for a few $10^8$ yr. 
Evolutionary synthesis modelling of post-starbursts in local massive gas-rich spiral --
spiral merger remnants have shown that these systems can also have
 tremendous bursts strengths that increased their stellar masses by $10-50$\%. In the case of NGC 7252, the very strong Balmer absorption lines can only be reproduced with a burst that increased the stellar mass by at least 30 and possibly up to 50\% (Fritze -- v. Alvensleben \& Gerhard 1994a, b). Starbursts in massive 
interacting galaxies hence are
completely off the burst strength -- galaxy mass relation for starbursts in
non-interacting dwarf galaxies, as seen in Fig. 1, raising the question if the SF process is the
same or different in normal SF regimes and dwarf galaxy starbursts on one hand
and in violent starbursts triggered by mergers of massive gas-rich galaxies on
the other (see also Fritze -- v. Alvensleben 1994).

\medskip\noindent
{\bf S}tar {\bf F}ormation {\bf E}fficiencies ({\bf SFE}s), as defined by the amount of stars formed out of a
given amount of gas ${\rm SFE := M_{stars}/M_{gas}}$ vary tremendously 
between the two regimes. On a global scale, SFEs are of order $0.1 - 5$ \% 
for spiral, irregular, and dwarf starburst systems (e.g. Murgia {\sl et al.} 2002), whereas they can reach $10 -
50$ \%  and $30 - 90$ \% in global and nuclear starbursts triggered by massive
gas-rich mergers like NGC 7252 and ULIRGs, respectively. Analysing the
post-starburst in NGC 7252 by means of evolutionary synthesis models GALEV, we
could show that for this galaxy even under the most conservative assumptions of
two particulraly luminous and particularly gas-rich progenitor spirals, the SFE
during the merger-induced starburst must have been very high, i.e. of order
SFE $\geq 40$ \% {\bf on a global scale}, i.e. 1 or 2 orders of magnitude higher
than in normal SF mode (Fritze -- v. Alvensleben \& Gerhard 1994b). SFEs as high as this had only
been thought to occur in the early universe. 

SFEs $\geq 20$ or even 50 \% are required for the formation of star clusters that are masive, compact, and
strongly bound enough to be able to survive for a Hubble time, i.e. for 
{\bf G}lobular {\bf C}lusters ({\bf GC}s), as shown in hydrodynamical 
simulations of star cluster formation
(Brown, Burkert \& Truran 1995, Elmegreen \& Efremov 1997, Li {\sl et al.}
2004). The high SFE found in NGC 7252 and
other mergers led to the idea that GCs might form in these events and not
exclusively in the early universe (Fritze -- v. Alvensleben \& Burkert 1995). By about the same time,
first HST images showed the rich star cluster systems now routinely found in
starbursts and mergers. The radii of these clusters remained ill-determined by that time due to the aberration problems of WFPC1.  

\medskip\noindent
While not much is known about molecular cloud {\bf structures} in interacting 
and merging galaxies, a clear difference does exist in the molecular gas
{\bf content} between normal SFing
galaxies and ULIRGS. This knowledge is based on the fact that while the CO(1-0) line traces molecular gas at densities
${\rm n \geq 100~cm^{-3}}$, the HCN(1-0) and CS (1-0) lines trace gas at
densities ${\rm n \geq 30\,000~cm^{-3}}$ and ${\rm n \geq 100\,000~cm^{-3}}$,
respectively. Sub-mm observations show that while for normal SFing 
galaxies only a small fraction $(\sim 0.1 - 3$ \%) of all 
their (CO) molecular gas is at the high densities of molecular cloud cores, as traced by HCN or CS, i.e.
${\rm L(HCN)/L(CO) \sim 0.001 - 0.03}$, this
very high density gas accounts for $50-90$ \% of all molecular gas in the central
regions of ULIRGs, for which ${\rm L(HCN,~CS)/L(CO) \leq 1}$. On scales 
of a few 100 pc, gas at molecular cloud core densities dominates the dynamical
mass in ULIRGs. The molecular cloud structure there can in no way resemble that
of the Milky Way and normal SFing galaxies with their tiny HCN cores within large
CO molecular clouds. Solomon {\sl et al.} (1992) and Gao \& Solomon (2004) find a very
tight correlation between SFRs derived from FIR luminosities and molecular cloud
core masses derived from HCN luminosities over a range of more than 3 orders of
magnitude in SFRs from normal spirals all through ULIRGs. They also show that,
albeit with considerable scatter, the ratio between L(FIR) and the mass of gas at
molecular cloud core densities is $\sim const$. This can be interpreted in
terms of SFRs directly proportional to the mass of gas at very high densities.

The famous Schmidt (1959) law relating the surface densities of SFR and HI by
${\rm \Sigma_{SFR} \sim \Sigma_{HI}^n}$ with ${\rm n \sim 1}$ for
spirals/irregulars and
${\rm n \sim 2}$ for ULIRGs (Kennicutt 1998), that is valid over 5 orders of
magnitude in gas surface density and 6 orders of magnitude in SFR density, 
becomes a universal ${\rm \Sigma_{SFR} \sim \Sigma_{HCN,~CS}^n}$ with ${\rm n
=1}$ for all SF regimes, when expressed in terms of high density molecular gas. 
Apparently, the gas at molecular cloud {\bf core} densities is transformed into stars
with almost 100\% efficiency on short timescales, and the efficiencies and
timescales for SF are set by the transformation of low density gas traced by
CO into high density gas traced by HCN or CS. This is an important issue to
consider in hydrodynamical modelling of galaxies and galaxy mergers which then
needs to account for a multi-phase ISM and include a careful description of
phase transitions, SF and feedback processes.  

SF in normal galaxies, spirals and irregulars, is thought to occur through the
collapse of molecular clouds, whereby the mass spectrum apparently remains self-similar
from molecular clouds through molecular cloud cores all the way to the mass
spectrum of open star clusters, all of which are power laws with ${\rm m \sim
-1.7~.~.~.~-2}$ (Lada \& Lada 2003, cf. Elmegreen \& Efremov 1997 for a 
theoretical foundation). 
In interacting galaxies, the frequency of molecular cloud collisions increases
strongly and this will considerably enhance SF. Moreover, molecular clouds get 
shock-compressed by external pressure (recently verified observationally for the Antennae galaxies by Haas {\sl et al.} 2005), grow denser and more massive, and this 
process can drive up the SFE very efficiently (Jog \& Solomon 1992, Barnes 2004). 
Jog \& Das (1992, 1996) have shown that a relatively small increase in the 
external ambient pressure to values 3 -- 4 times the internal pressure within the
molecular clouds in the undisturbed galaxy can drive SFEs up to 70 -- 90 \%. 

A first attempt to assess the molecular cloud mass spectrum in the nearest
ongoing merger NGC 4038/39 by Wilson {\sl et al.} (2003) revealed a power law with 
${\rm m \sim-1.2~.~.~.~-1.6}$ but remained limited to a mass range above ${\rm
10^7~M_{\odot}}$. Resolution of molecular clouds below that and observations of
molecular cloud cores have to await ALMA,
and the same is true before we can know if the molecular cloud mass spectrum is
different or not in massive gas-rich mergers from what it is in non-interacting
galaxies. ULIRGs in any case show that, averaged over volumes of $10 - 300$ pc, the
ratio M(HCN)/M(CO) can reach up to $0.3 - 1$, i.e. that the molecular cloud
structure is very different indeed -- to the point that it becomes very difficult to
imagine much internal structuring at all, if essentially all the molecular gas
is at molecular cloud core densities. 

\section{Star Cluster Formation}
The Milky Way, M31, LMC, SMC,... all are forming open clusters with masses ${\rm
\sim 10^3~M_{\odot}}$, low concentration, and a power law cluster mass spectrum. 
With their short lifetimes $\sim 10^8$ yr, these open clusters will soon dissolve into the
field star population. All these galaxies also have Globular Clusters ({\bf
GC}s) with high masses ${\rm \sim 10^{5.5}~M_{\odot}}$, high concentration, a
Gaussian GC mass spectrum and lifetimes of order a Hubble time. 
The LMC features an intriguing gap in star cluster ages with only one cluster
in the age range between 4 and 13 Gyr, although field star formation and
chemical enrichment proceeded continuously. Star cluster formation seems to
only or predominatly have occurred in epochs of enhanced field star formation
that can be associated with close passages of the SMC and/or the Milky Way (cf.
Rich {\sl et al.} 2001). Similarly, star cluster formation in M51 is found 
to have been
significantly enhanced during the last close encounter with its companion NGC
5196 (Bastian {\sl et al.} 2005). With SFEs in the normal range, the ongoing
cluster formation in normal, non-interacting as well as in dwarf starburst
galaxies is expected to produce open clusters rather than GCs. If some locally
exceptionally high SFE might produce a GC is an open issue. The fact that no or
at most very few intermediate age GCs are known in normal galaxies confirms that
this is at best a very rare case. 

Larsen (2004) reports the detection of so-called {\bf S}uper {\bf S}tar {\bf C}lusters ({\bf
SSC}s) in a number of undisturbed normally SFing face-on spirals. 
These SSCs are clearly very bright and 
very young, at least some of them have been shown to be very massive, too
(Larsen {\sl et al.} 2004). With
masses around ${\rm 10^{5-6}~ M_{\odot}}$ and small radii, they resemble young
GCs, although GC formation is not expected in these probably normal, i.e. low
SFE environments. If they really were young GCs forming in a non-spectacular way during
normal SF in undisturbed spirals, however, we might ask: ``Where are the 
descendants of all those SSCs that formed earlier-on, i.e. where are all the
intermediate-age GCs in those spirals? Or are we whitnessing a very special
epoch in the life of those spirals? And, in which sense is it special?''

\subsection{Star Cluster Masses}
There are two fundamentally different methods to assess the masses of {\bf Y}oung 
{\bf S}tar {\bf C}lusters ({\bf YSC}s). Dynamical mass estimates on the basis of
central stellar velocity dispersions yield results independent of any assumption
about the stellar IMF. Requiring spectroscopy, however, this method is
time-consuming and limited to the brightest and nearest systems. Mass 
segregation will lead to systematically underestimate dynamical masses. It has
been shown to not only occur secularly in the course of dynamical evolution, but
to some part already to be built in at birth for YSCs in the LMC by de Grijs
{\sl et al.} (2002a, b). Spectroscopy of the brightest SSCs in nearby undisturbed face-on
spirals embarrassingly indicates masses in the range ${\rm 10^{5-6}~ M_{\odot}}$ 
for those objects with half-light radii similar to those of GCs (Larsen {\sl et
al.} 2004). 

\noindent
Photometric mass estimates, on the other hand, use multi-$\lambda$ photometry
in combination with a grid of evolutionary synthesis models like our GALEV
models for {\bf S}imple 
{\bf S}tellar {\bf P}opulations ({\bf SSP}s with all stars of the same age and
metallicity) like star clusters and a dedicated SED Analysis Tool to
derive ages, metallicities, extinction values, and masses, including their 
respective 1$~\sigma$ ranges {\bf for all
clusters in the field} (Anders {\sl et al.} 2004a, Anders \& Fritze -- v. Alvensleben, {\sl these proceedings}). 4 reasonably 
chosen passbands (e.g. U, B, V or I, and a
NIR band) are enough to obtain reasonably precise estimates of all the relevant
parameters, including masses. Based on 4-band imaging, these photometric mass
estimates are economic and far-reaching, and yield parameters for all clusters
in a field in 4 shots. They have to assume a stellar IMF, and the accuracy of the stellar
masses derived strongly depends on the precision of the age determination due to the steep time evolution of
M/L-ratios in early stages. Being based on photometric magnitudes, the precision of photometric mass estimates strongly depends on well-defined cluster radii and accurate 
aperture corrections (Anders \& Fritze -- v. Alvensleben {\sl this volume} and 
Anders {\sl et al., submitted}). 

\subsection{YSC Formation} 
YSC formation is observed in a wide variety of environments. E.g., in addition to
the 3 well-known SSCs in the isolated dwarf starburst galaxy NGC 1569,
Anders {\sl et al.} (2004b) identified 166 YSCs on HST archival 
images, more than 1000 YSCs are seen in the ongoing merger between NGC 4038 and
4039, the famous Antennae galaxies (Whitmore \& Schweizer 1995, Whitmore 
{\sl et al.} 1999, Fritze -- v. Alvensleben 1998, 1999, Anders 
{\sl et al., in prep.}). YSCs are seen in many
other interacting and/or starburst galaxies. Sometimes they are found all over the main body of
an interacting system, as in the Antennae or NGC 7252, sometimes they are
confined in or around a starburst nucleus as is the ULIRGs Arp 220 and NGC 6240
(cf. Shioya {\sl et al.} 2001, Pasquali {\sl et al.} 2003). YSCs are also found all along
some, but not all, extended tidal features, like e.g. all along the 120 kpc 
tidal tails of the Tadpole and Mice galaxies shown in Fig. 2 (Knierman {\sl et al.} 2003, de Grijs {\sl et al.} 2003a), as
well as in group environments like Stephan's quintet (cf. Gallagher {\sl et al.} 2001). 

\begin{figure}
\begin{center}
\psfig{figure=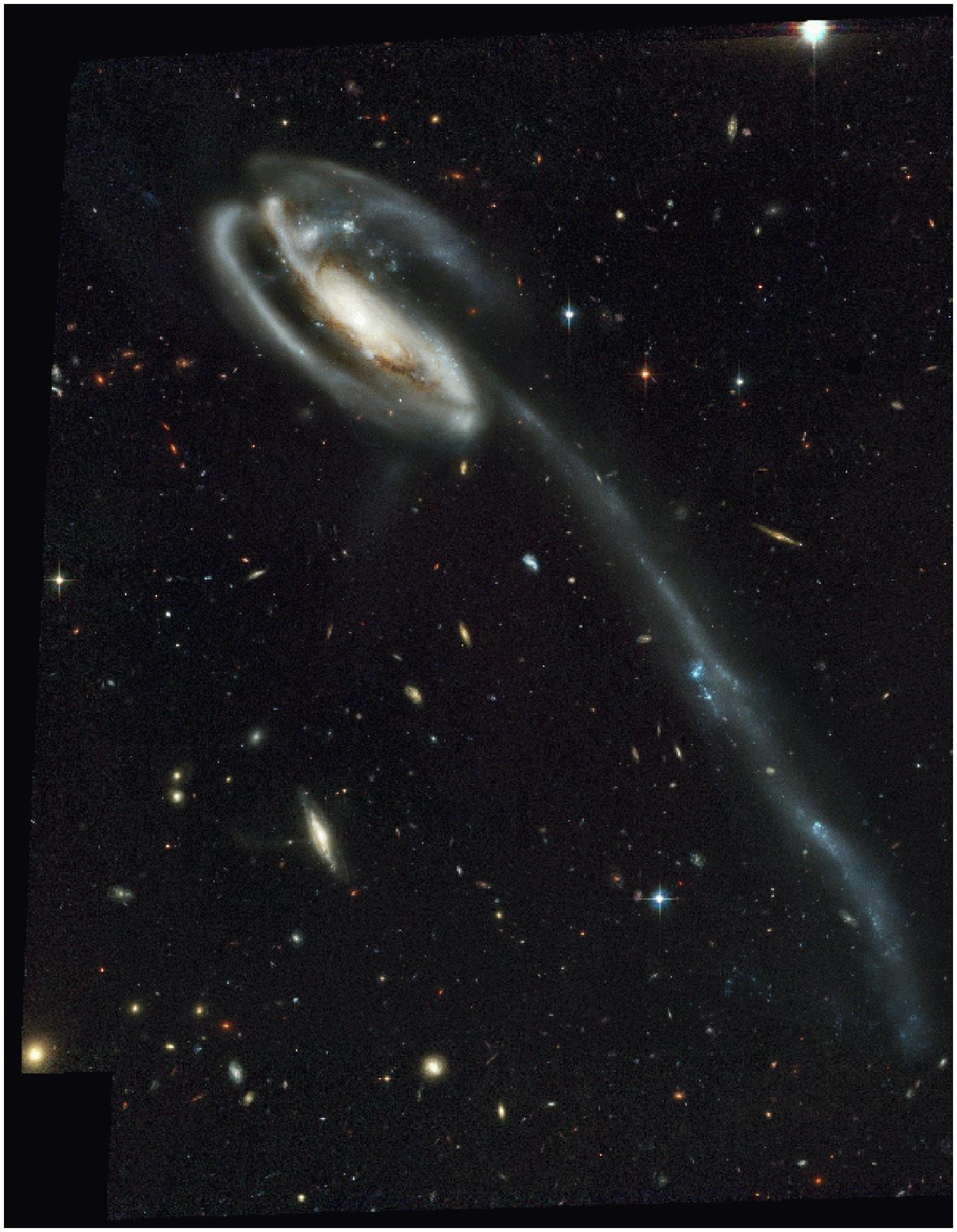,height=6.cm}\hspace{1.cm}\psfig{figure=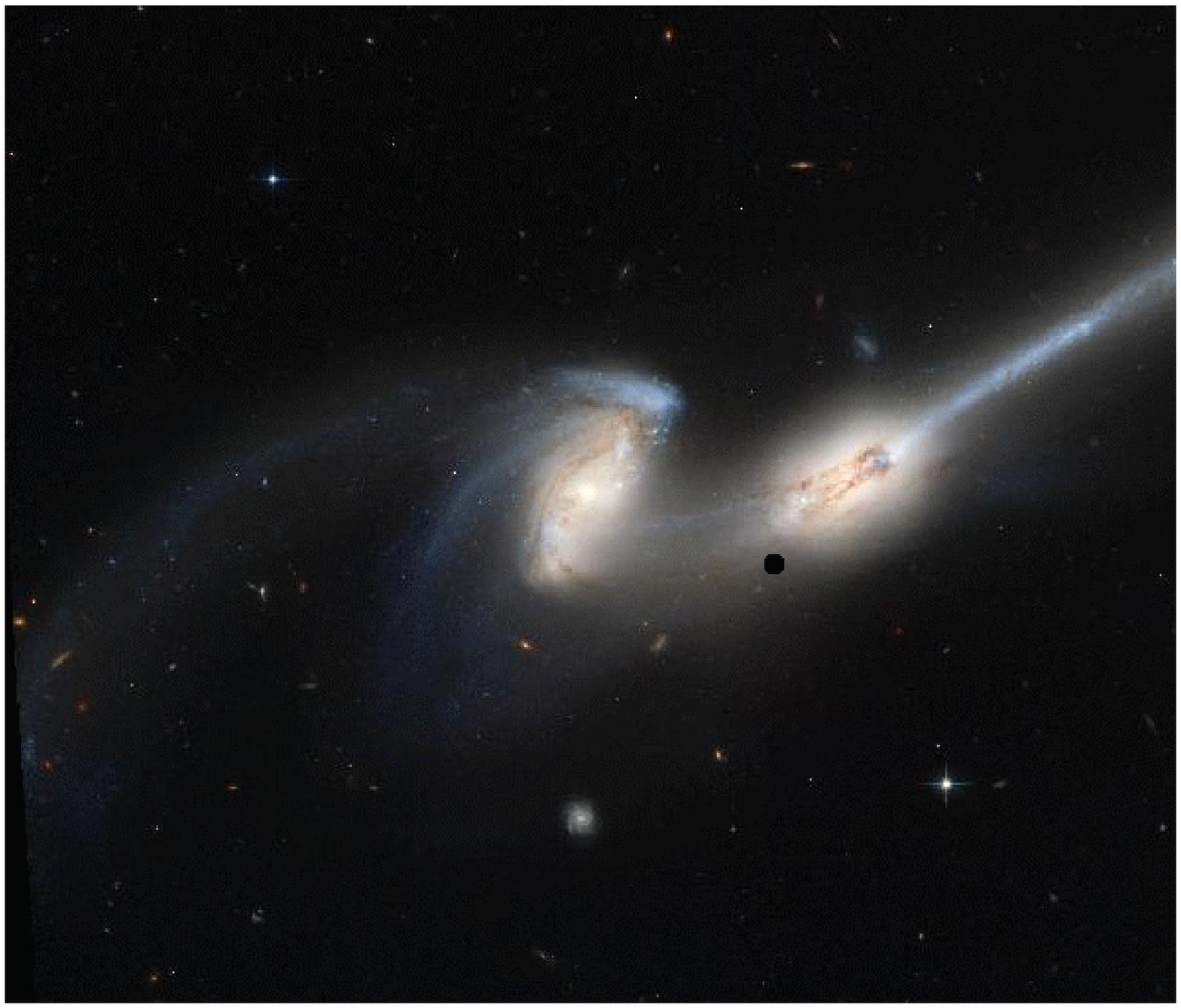,height=6.cm}
\end{center}
\caption{ACS Early Release Observation of the Tadpole (left) and Mice galaxies (right) showing large
numbers of young blue star clusters all along their $\sim 120$ kpc long tidal tails (Credit: NASA, H. Ford (JHU), G. Illingworth (UCSC/LO), M.Clampin (STScI), G. Hartig (STScI), the ACS Science Team, and ESA).}
\end{figure}

These environments cover all the range from very dense regions within or close
to an active starburst nucleus out to expanding low-surface brightness and
probably also low physical density regions far from the galaxy centers and 
we may ask
the question whether or not the YSCs in these very different environments are
similar -- individually or as a population. In the Milky Way and other local
galaxies we are used to distinguish between open clusters on the one hand and GCs
({\bf GC}s) on the other, with open clusters being low-mass (${\rm \leq
10^3~M_{\odot}}$) clusters with low
central concentration, described by Elson, Fall \& Freeman (1987) 
surface brightness 
profiles, weakly bound, short-lived with lifetimes of order a few $10^8$ yr, and, hence, predominantly young. Only very
few open clusters in the Milky Way have ages of 1 Gyr and beyond, those reside
in peaceful isolation very far from the Galactic Center. GCs, on the other hand,
have high masses ${\rm \sim 10^{5.5}~M_{\odot}}$, high central concentrations,
are described by tidally truncated King models in their surface brightness
structure, they are strongly bound and, hence, long-lived, have ages of
order 13 Gyr. We do not know whether open clusters and GCs are two different
types of objects -- by nature or by nurture -- or whether they rather form two
ends of one continuous distribution in terms of mass and/or concentration. Nor
do we know whether or not the formation processes leading to open and globular
clusters, respectively, are the same or different, if the formation of one of
both types preferentially or exclusively occurs in specific environments or if
always the full spectrum of clusters from very massive and strongly bound to
very losely bound low-mass clusters is formed. 

The young open cluster populations in the Milky Way, M31, LMC, SMC, and other nearby non-interacting,
peacefully SFing galaxies feature power-law luminosity functions, while the old
GC populations in these galaxies show Gaussian type luminosity functions with 
${\rm \langle M_V \rangle =-7.3}$ mag, ${\rm \sigma(M_V)=1.3~mag}$ and Gaussian 
type mass functions with ${\rm Log \langle M_{GC} /M_{\odot} \rangle =5.3}$. 

We do not know much yet about open and globular clusters in starbursting and interacting
galaxies, in particular because of the difficulty to distinguish between both
types in a YSC system. What GCs and a GC system looked like when they were 
young is another open issue. For the Milky Way GC system it is for sure 
that what is left today from the original GC system is just the {\sl hardiest
survivors of an originally much larger population}, as stated by W. Harris.

The luminosity of the most luminous
YSC apparently increases with increasing number of YSCs in more actively starforming
galaxies, as predicted in numerical simulations by Kravtsov \& Gnedin (2005). 
It is not clear yet, if this increase is only what is expected from a
purely statistical size of sample effect or if there is a systematic effect 
on top of that. This kind of study in any case has to look at cluster masses 
instead of luminosities, e.g. on the basis of 4 passband photometry in the 
way described above, because of the very rapid fading of YSCs in early
stages. Our models have shown that e.g. at half-solar metallicity a YSC 
fades by
5 mag in V over the first Gyr with 3.3 mag of fading already in the first 
$10^8$ yr due to stellar evolution only (Schulz {\sl et al.} 2002, 
Anders \& Fritze -- v. Alvensleben 2003). If a YSC also loses stars (and how
many) by already 
dissolving into the field star population depends on its initial parameters mass, density, and velocity dispersion (cf. Bastian {\sl et al.} 2005, Lamers {\sl et al.} 2005, Takahashi \& Portegies Zwart 2000, Spitzer 1987). 

\subsection{YSC Masses and Mass Functions}
The rapid evolution of cluster luminosities in early stages is the reason why
luminosity functions of YSC systems need not reflect the shape of their
underlying mass functions. If the age spread among the clusters is of the order
of the cluster ages, as is the case for YSC systems forming in massive gas-rich
spirals 
with typical burst durations of a few $10^8$ yr, the luminosity function may
well look like a power-law with the mass function showing a clear turn-over like
the Gaussian mass function of old GC systems (Meurer 1995, Fritze -- v.
Alvensleben 1998, 1999). Already when analysing the WFPC1 data in U, V, and I 
obtained by 
Whitmore \& Schweizer (1995) for the Antennae, assuming half-solar metallicities
 and carefully accounting for the
individual ages of the 393 YSCs with precise enough color information, we 
found a Gaussian cluster mass spectrum with ${\rm log \langle M_{YSC}/M_{\odot} \rangle \sim 5.6}$ and $\sigma = 0.46$, very similar to the GC mass spectra in the 
Milky Way, M31, and nearby elliptical galaxies. The major drawback in our analysis was our
assumption of a uniform reddening for all YSCs lack of more detailed
information about individual clusters. Zhang \& Fall (1999) used reddening-free
Q-parameters in their analysis of the deeper WFPC2 data and found a power-law mass
function. The major drawback in their analysis was that they had to exclude an 
important fraction of clusters for which the Q-parameters did not yield an
unambiguous age. Excluding this age group of clusters in our models also leads
to a power-law MF. Hence, by that time, the MF of YSCs forming in
this merger-induced starburst remained controversial. A reanalysis of WFPC2 data including H$_{\alpha}$ by Fall (2004) confirmed their earlier result of a power-law MF for the YSCs (age $<10^8$ yr) in the Antennae. 
Recently, we have begun to 
reanalyse the
WFPC2 data for the Antennae with our SED Analysis Tool and very conservatively 
identify $\sim 1000$ clearly extended YSCs with photometry in UBVI accurate 
to at least 0.2 mag. We are currently examining very carefully all sources of
uncertainties. Among others, we account for different completeness limits for
clusters in regions of different background and apply appropriate aperture
corrections individually for each cluster. We thus obtain for each cluster 
its individual metallicity,
extinction, age and mass and we will therefore be able to 
determine with unprecedented significance 
whether or not the YSC luminosity function will look like a power law as for open clusters in the 
Milky Way or already show a turn-over like the luminosity function of old
GC systems and the luminosity functions of intermediate-age star cluster systems in NGC
1316 (Goudfrooij {\sl et al.} 2004) or the star clusters in the post-starburst
region M82-B (de Grijs {\sl et al.} 2003b) (cf. Anders {\sl et al., in prep.}). 

\subsection{Evolution of Star Cluster Populations}
Beyond the evolution of each individual cluster due to stellar evolution and the
 mass loss it brings along, star clusters also lose mass in terms of stars in the course of their dynamical evolution. Stars are lost from the tail of the Maxwellian stellar
 velocity dispersion, that, in turn, keeps being replenished due to 2-body relaxation.
 This effect leads to the evaporation of star clusters and, of course,
destroys low-mass clusters in particular. This evaporation due to 2-body
 relaxation of stars within the cluster can be enhanced by tidal shocking if a
 cluster on its orbit sees a variable potential, e.g. by crossing a disk or by
 passing close to a galaxy center on an excentric orbit. It may then also
 destroy higher mass clusters on short timescales. These effects are included
in semianalytical and numerical dynamical modelling of star cluster 
populations in Galactic potentials, timescales depend on the mass of a cluster
and on its initial concentration (cf. Chernoff \& Weinberg 1990, Fall \& Zhang
2001). Both the initial cluster mass function and the initial distribution of
densities or concentrations are not known for the progenitors of old GCs, nor for YSC systems
in interacting and starburst galaxies. 

\noindent
Another process that can only be 
accounted for in numerical models that consistently include the internal 
dynamics within clusters evolving in a Galactic potential is dynamical 
friction affecting predominantly high-mass clusters near the galaxy centers.
Both for a Milky Way-type and an elliptical galaxy potential, Vesperini and
collaborators find that the destruction of low-mass clusters by evaporation and the
destruction of high-mass clusters by dynamical friction largely balance each other
and that despite of the destruction of more than 50 \% of the original cluster
population in the course of evolution an initially Gaussian-shaped cluster mass
function is generally preserved while an initially power-law type mass function requires
significant fine-tuning of all the parameters involved in the modelling to be
transformed into the Gaussian observed for Milky Way GCs (Vesperini \& Heggie
1997, Vesperini 2000, 2001, Vesperini \& Zepf 2003). All this, however, has only been
calculated for static galaxy potentials. In the time-varying potential of an
ongoing merger like the Antennae, where violent relaxation, external pressure, 
violent SF and feedback are
vigorously at work, things are definitely more complicated. 

Timescales for star cluster destruction processes can also be derived from
observations of YSC systems. Under the assumption that YSCs were formed at a
continuous rate and with similar properties over the duration of a starburst, 
comparison of YSC subgroups
from several age bins within the typically $2-4 \cdot 10^8$ yr duration of
starbursts in massive gas-rich mergers, can show what fraction of clusters still
seen in the youngest bin is already missing from the older bins. This way,
timescales for the fastest cluster destruction processes can be estimated
empirically and are found to be in broad agreement with theoretical estimates
(cf. Boutloukos \& Lamers 2003, Lamers {\sl et al.} 2005). If the assumptions of
a constant YSC formation rate during the burst and of uniform YSC properties are correct or not is crucial
for this approach and remains to be tested by detailed observations of YSC
systems and subpopulations in various environments. 

In any case, it is crucial to account for destruction effects when
comparing star cluster (sub-) systems of various ages and deriving initial
cluster mass functions. At present, it remains to be seen if and in how far
initial mass functions of YSC systems are universal or dependent on environment
and if they resemble initial GC mass functions or not. As long as we cannot resolve molecular clouds and molecular cloud cores in interacting systems and determine their mass spectra, the initial mass function of the YSCs forming in these systems has to serve us as a proxy and is the only way to check in how far the SF process is the same or not in the strong merger-induced starbursts and in normal SF regimes.   

The masses of the YSCs in the Antennae range from a few 100 ${\rm M_{\odot}}$ for
the smallest OB association-like clusters through  ${\rm \geq 10^6~M_{\odot}}$ (Mengel {\sl et al.} 2002), i.e. 
definitely into the range of GC
masses. Their radii, as determined by
Whitmore {\sl et al.} (1999), are within the range of Milky Way GC radii, a
more careful redetermination by P. Anders is under way, as well as
correspondingly improved photometry, photometric masses, and their mass function. 

NGC 1569, on the other hand, is an isolated dwarf galaxy 
with a strong starburst going on. Three so-called Super Star Clusters and 
some 44 YSCs were previously known (Hunter {\sl et al.} 2000). On HST archival images provided to
us by an ESO ASTROVIRTEL project (PI R. de Grijs) Anders {\sl et al.} (2004b) 
identified 
169 YSCs in total with accurate photometry in U, B, V, I, and H 
(errors $< 0.2$ mag), for which by means of our SED Analysis Tool 
individual metallicities,
ages, extinction values, and masses including their respective $1\sigma$ errors
could be determined. The YSCs in this system turn out to have masses in the
range ${\rm 10^3~-~10^4~M_{\odot}}$, only three clusters have masses beyond a few
${\rm 10^5~M_{\odot}}$, the average mass of old GCs. Hence, we conclude from the
masses alone that no or, at most, very few young GCs are formed in this 
non-interacting dwarf starburst galaxy. 

\subsection{The Fate of Young Star Clusters in Mergers}
From the example of NGC 7252, where a large number of star clusters that formed in
the burst between 650 and 900 Myr ago are still alive and compact, we know that
GCs indeed did form in this massive gas-rich spiral -- spiral merger, since those clusters have already survived many crossing times. In this respect, NGC 7252, one of the most widely (from X-rays
through radio) observed merger remnants, for which we had found the very high SFE
in the first place, holds a key role in that it really shows that a number of
clusters with GC masses and radii do still exist at ages where most open
clusters already were destroyed, and in a number comparable to the number of GCs typically inherited from the two spirals. 
What we do not know, on the other hand, is how many YSCs were
formed in total in NGC 7252, i.e. what fraction of the originally formed entire
YSC population did survive and hence, merit to be called young GCs. 

\medskip\noindent
We are unable to tell apart the YSC populations in the Antennae or NGC
1569 individually into young open clusters and young GCs. As we have shown
before, neither their luminosities nor their radii fall into two distinct
categories. At their ages, these YSCs are not expected to be tidally truncated
yet by their parent galaxy potential and Whitmore {\sl et al.} (1999) show that the YSCs \#405 and \#430 in the Antennae show
HST determined surface brightness profiles without indication of a tidal
truncation, similar to the young (8 -- 300 Myr) clusters in the LMC (Elson, Fall \& Freeman 1987). Hence, the classical measure of concentration parameters ${\rm c:=
log (R_{tidal}/R_{core})}$ cannot be applied. 

We are trying hard (Anders {\sl et al., in prep.}) to see any
difference or gap in YSC properties in the Antennae in terms of
masses, half-light radii, or densities after our very careful radius determinations. Very preliminary results fail to indicate any significant differences, neither within the entire population
nor between different age groups, in agreement with Elmegreen \& Efremov's (1997) theoretical scenario. They suggest that supersonic turbulence produces a scale-free fractal structure in the gas and a universal m $\sim -2$ power-law mass spectrum for the clouds and for the forming clusters. Cloud mass and ambient pressure determine the degree of internal binding in the nascent cluster and, hence, if the result is either unbound, or a weakly bound open cluster, or a strongly bound GC. In their scenario, by raising the ambient pressure, galaxy interactions can lead to a mode of SF in which massive, dense, and tightly bound clusters are the {\bf primary} result, although to some degree the entire continuum down to losely bound/unbound clusters should also be present.

\medskip\noindent
We know both from dynamical modelling and from evolutionary synthesis 
that spiral  -- spiral mergers can well evolve into E/S0 (and in some cases 
even Sa) galaxies morphologically and spectroscopically (e.g. Barnes \&
Hernquist 1992, Bournaud, Combes \& Jog 2004, Springel \& Hernquist 2005, Fritze -- v. Alvensleben \& Gerhard 1994a). The spectroscopic type of the merger remnant
depends on whether or not the SFR after the merger-induced starburst goes to
zero or remains at some finite value. Here again, NGC 7252 holds a key role:
Hibbard {\sl et al.} (1994) have observed the amount and kinematics of HI along its
tidal tails and find that there is a considerable reservoir of HI falling back
onto the body of the merger remnant on long timescales of order $3-4$ Gyr. In
simulations by Hibbard \& Mihos (1995) this gas is seen to settle into a radially
growing HI disk. We predict a comparable amount of gas to be 
released by dying burst stars (Fritze -- v. Alvensleben \& Gerhard 1994b). All this gas can be transformed into a stellar 
disk if the present SFR of ${\rm 1.5 - 3~M_{\odot}~yr^{-1}}$ would continue in NGC 7252.

\medskip\noindent
We hence expect that those ellipticals, S0s (and
Sa's), that are remnants from massive gas-rich mergers should feature a
younger and more metal-rich subpopulation among their GCs with the age of this
subpopulation dating back to when the merger occured and the metallicity being
roughly set by the ISM abundance in the merging spirals. I put ``roughly''
because of metallicity gradients within the spirals, differences between various
spiral types, and the possibility that clusters formed late in the burst could
already be burst-enriched, an effect that would reveal itself by an
$\alpha$-enhancement in the cluster spectra, as tentatively seen in the spectrum of the YSC W3 in
NGC 7252 (Schweizer \& Seitzer 1993 and Fritze -- v. Alvensleben \& Burkert 1995). 

In Fritze -- v. Alvensleben (2004) I have used GALEV models for the chemical
evolution of various spiral types to show that despite considerable scatter a
broad age -- metallicity relation exists for spiral galaxies, analogous to 
the one for
stars in the Milky Way. The later the merger occurs, the higher will be the
metallicity of the star clusters expected to form. 

\noindent
Zepf \& Ashman (1993) found that the specific GC
frequency, i.e. the number of GCs per unit galaxy mass, is typically twice as
high in E/S0s as in spirals and they predicted that an average elliptical can 
only result
from a merger of two average spirals if a number of GCs can form in the burst
that is of the order of the number of GCs present in the two spirals before
they merge. This has indeed been the case in NGC 7252, as we showed in Fritze
-- v. Alvensleben \& Burkert (1995). 

\noindent
A number of intermediate age GC systems have been reported in merger remnants and
dynamically young ellipticals (e.g. Goudfrooij {\sl et al.} 2001a, b). And many, if
not most, bright ellipticals and S0s do feature bimodal color distributions for
their GC systems in terms of $V-I$ (Gebhard \& Kissler-Patig 1999, Kundu \& Whitmore 2001a, b). 
The blue peak seems to be fairly universal and similar to that of the Milky Way
halo GC population, the position and relative height of the red GC color peak is
variable. 
The interpretation of these bimodal GC color distributions, however, is not
straightforeward due to the well-known fact that optical colors are degenerate
in terms of metallicity and age. A young and metal-rich stellar population can have the
same color as an older and metal-poor stellar population (e.g. Worthey 1994). 
While the merger scenario predicts two
GC subpopulations, a hierarchical formation scenario would be expected to produce
a broad or multi-peaked GC color distribution. Other scenarios for the {\sl in
situ} formation of a second GC subpopulation have also been proposed
(e.g. Forbes \& Forte 2001). 
In Fritze -- v. Alvensleben (2004) I have shown with help of our GALEV models
for the spectrophotometric evolution of star clusters, how -- depending on its
initial metallicity -- the color distribution of a secondary GC population evolves
with time. A low metallicity cluster population with [Fe/H]$=-1.7$ similar to
the Milky Way halo GCs would have its ${\rm \langle V-I \rangle}$ at $\sim 0.6$ at an age of 
300 Myr and move towards the universal blue ${\rm \langle V-I \rangle}$ 
peak around 0.9 by an age of
12 Gyr. If a star cluster population starts out with a metallicity around
[Fe/H]$\sim -0.4$, it would attain ${\rm \langle V-I \rangle} \sim 1.2$ 
around 12 Gyr, similar to the red-peak GCs in several E/S0s. Because of the age
-- metallicity degeneracy a red peak at ${\rm \langle V-I \rangle} \sim 1.2$ can
in principle result from a manifold of very different combinations of age and
metallicity, ranging from very young and very metal-rich (2 Gyr, [Fe/H]$=+0.4$)
all through very old and metal-poor (15 Gyr, [Fe/H]$=-1.7$). Individual GC
spectroscopy is feasible with 10m class telescopes out to distances
of $\sim 20$ Mpc and allows to disentangle ages and metallicities by measuring 
Lick indices. It will, however, remain very time-consuming and restricted to the
brightest GCs in these distant galaxies. HST imaging will be required in
order to secure that spectra indeed refer to individual clusters, not to blends. 
T. Lilly is currently developing an Analysis Tool for Lick spectral indices  in
terms of ages and metallicities including their 1$\sigma$ uncertainties in
analogy to the SED Analysis Tool developed by P. Anders (Lilly \& Fritze -- v.
Alvensleben 2005a {\sl submitted}, 2005b {\sl in prep.}).
Multi-band imaging will, however, remain the most powerful tool for the analysis of
significant fractions of GC populations in external galaxies and I could also 
show that already the inclusion of photometry in one additional NIR passband can
to a fairly large extent resolve the age -- metallicity degeneracy. E.g. will
the two GC systems mentioned above with different age -- metallicity
combinations yielding the same ${\rm \langle V-I \rangle} \sim 1.2$ show well
distinguishable ${\rm V-K}$ colors: ${\rm \langle V-K \rangle} = 3.5$ for 
(2 Gyr, [Fe/H]$=+0.4$) and ${\rm \langle V-K \rangle} = 2.3$ for 
(15 Gyr, [Fe/H]$=-1.7$).

\section{Star Cluster Formation and Star Formation}
Star cluster formation clearly is an important mode of SF in starbursts. Meurer
{\sl et al.} (1995) already pointed out that $\sim 20$ \% of the UV-light in
starbursts is from star clusters, they tentatively conclude from a sample of
starburst galaxies that the UV-light contribution from clusters
relative to the total UV-light seems to increase with intrinsic UV surface
brightness of the galaxy. Elmegreen \& Efremov (1997) and Kravtsov \& Gnedin (2005) expect from numerical simulations that the fraction of SF that goes into star cluster formation and into massive, dense, and tightly bound young GCs in particular, should increase in high external pressure and, hence, high SF environments. 

\noindent
In our pixel-by-pixel analysis of the HST ACS Early
Release Observations of the Tadpole and Mice galaxies we found large numbers of 
YSCs with characteristic masses of $\sim 3 \times 10^6 {\rm M_{\odot}}$ 
all along
the tremendous 120 kpc long tail of the Tadpole and along the prominent tail of
one of the Mice galaxies and we estimated their light
contribution to amount to $\sim 70$ \% in the B-band and to $\sim 40$ \% in I
(de Grijs {\sl et al.} 2003a). We concluded that both in the Mice and the Tadpole
galaxies more than 35 \% of all SF along the extended tidal structures went 
into star cluster formation, even into the formation of massive clusters
with characteristic masses in the range of GC masses. In view of the expanding nature and presumably low physical density of the low surface brightness tails this appears surprising. SCs are very young, $(1.5-2) \cdot 10^8$ yr, profiles and radii are hard to come by (diameters $\leq 35$ pc), so it is difficult to estimate for how long these YSCs are expected to survive. 

\section{Summary, Open Questions, and Outlook} 
Stars and SCs form from molecular clouds with the timescale and efficiency of SF being set by the transformation of lower density molecular gas, as traced by CO, into high density gas as traced by HCN- and CS-lines. Shocks and external pressure as occuring in galaxy mergers can greatly enhance this transformation,  increase the SF efficiency by up to two orders of magnitude, and possibly the ratio of SF that goes into SC formation as well as the degree of internal binding of the emergent clusters. We have no information yet about the molecular cloud structure in the high pressre environments caused by mergers but we know from integrated CO-, HCN-, and CS-luminosities that in ULIRGs, all of which are advanced stages of massive gas-rich mergers, essentially all of the molecular gas ($50 - 100$ \%) is at the very high densities typical for molecular cloud cores. It seems difficult to envisage much structuring within this gas. 

\noindent
With ALMA, it will be of prime interest to compare the molecular cloud structure
and mass spectra in interacting galaxies with those of galaxies undergoing
normal SF to check whether or not the SF process itself is universal with a
tremendous dynamical range or whether there are intrinsic differences between
normal and violent SF modes. As long as we cannot yet resolve molecular clouds and cloud cores in interacting galaxies, the YSCs forming in these systems will have to serve as a proxy.

\noindent 
It is clear that star cluster formation is an important mode of SF -- already in
the Magellanic Clouds, in actively SFing undisturbed spirals with SSCs, and, in
particular, in the strong starbursts triggered by interactions and merging
between gas-rich galaxies. 

\noindent 
Not yet clear on the other hand is whether the relative amount of SF that goes
into star cluster formation increases with increasing SFR, burst strength b, SFE, etc., as expected from theoretical models.

\noindent
It is clear that the formation of strongly bound and hence long-lived star
clusters like GCs requires exceptionally high SFEs that apparently were
ubiquitous in the early universe and are still achieved occasionally on
galaxy-wide scales in violent
gas-rich mergers like NGC 7252 that did form systems of new GCs comparable in richness to the
preexisting ones. 

\noindent
Not clear is whether there is a threshold in SFR, b, or SFE, below which GCs cannot
be formed at all or if SFEs high enough for GC formation can in some cases be
reached very locally within an undisturbed spiral, giving birth to just one or
few GCs. 
A careful analysis of spiral galaxy GC populations with respect to possible
outlyers in terms of age and metallicity should tell. This analysis, however,
gets complicated by the fact that such younger and more metal-rich GCs will not
stand out in optical broad-band colors due to the age -- metallicity degeneracy.

\noindent
It is not clear yet if the maximum star cluster mass scales with SFR, b, SFE,
and/or total number of YSCs. It is not clear either if in strong merger-induced
starbursts like the Antennae or NGC 7252 the usual weakly bound open clusters
and associations also form and to what extent. I.e., it is not clear if YSCs are
the same or different -- individually or as a population -- in different
environments like isolated and interacting, dwarf and giant galaxies, within
their central parts and far out along expanding tidal structures. The
starburst in the isolated dwarf galaxy NGC 1569 apparently did not form any GCs,
or at maximum very few, as deduced alone from the masses of its numerous and
bright YSCs. Gas-rich dwarf
-- dwarf galaxy mergers, if they weren't so hard to find, seem to be a very interesting case to test if they can
reach the very high SFE regimes, if it is the short dynamical timescales, the
shallow potentials of dwarf galaxies or the lack of ambient pressure that
accounts for the low SFEs in the isolated dwarf starbursts studied so far. 
 
\noindent
We still are unable to tell which and how many of the YSCs are open clusters and
which and how many are young GCs. Their masses, radii, and mass function seem
to be the key issues. Comparison of subpopulations of various ages in
mergers/starbursts going on for a while may hold clues, provided SC destruction effects are taken into account. 
A comparison between the relative amounts of SF that go into fields stars and
short-lived clusters on the one hand and into strongly bound and long-lived
clusters on the other over a wide range of bursts strengths, SFEs and
environments (non-interacting vs. mergers, gas-rich vs. gas-poor, giant vs. dwarf) would greatly enhance our
understanding of global SF in galaxies. 

\noindent
Long-lived star clusters, in any case, are valuable tracers of the (violent) SF
histories of their parent galaxies -- much better suited than integrated light,
because they can be analysed one-by-one, are easy to interprete (one age, one
metallicity), and the age and metallicity distributions of star cluster systems
give direct insights into their parent galaxies' formation and chemical enrichment
history. Accurate 4-passband photometry over a long enough wavelength basis,
including U (or at least B) and one NIR band, in combination with a dedicated
analysis tool can largely disentangle the age -- metallicity degeneracy and
allows to simultaneously determine individual ages, metallicities, extinction values, and
masses including their respective 1$\sigma$ uncertainties for large cluster
samples. Resolution of the clusters and accurate cluster radii are the critical
issues. 

\noindent
Star clusters are of threefold importance: they are uniquely suited to study the very SF and SC formation processes and their universality vs. environmental dependence, to study their parent galaxy's formation, evolution and chemical enrichment histories, and, as long as we cannot even resolve molecular clouds and cloud cores in the nearest interacting galaxies, the youngest SC systems  have to serve as a proxy for the molecular cloud structures they were born from.  
\section*{Acknowledgments} I gratefully acknowledge travel support from the organisers without which I could not have attended this conference and I thank my collaborators on this subject Richard de Grijs, Peter Anders and Thomas Lilly for their important contributions to my understanding.

\end{document}